\documentclass[superscriptaddress,showpacs,preprintnumbers,amsmath,amssymb,twocolumn,aps]{revtex4-1}

\usepackage{graphicx}
\usepackage{dcolumn}
\usepackage{bm}
\usepackage{verbatim}
\usepackage[usenames, dvipsnames]{xcolor}
\usepackage{float}
\usepackage{upgreek}

\def\be{\begin{equation}}
\def\ee{\end{equation}}
\def\bea{\begin{eqnarray}}
\def\eea{\end{eqnarray}}

\begin{document}

\title{Long coherent dynamics of localized excitons in (In,Ga)N/GaN quantum wells}

\author{S.~V.~Poltavtsev}
\email{sergei.poltavtcev@tu-dortmund.de}
\affiliation{Experimentelle Physik 2, Technische Universit\"at Dortmund, 44221 Dortmund, Germany}
\affiliation{Spin Optics Laboratory, St.~Petersburg State University, 198504 St.~Petersburg, Russia}
\author{I.~A.~Solovev}
\affiliation{Experimentelle Physik 2, Technische Universit\"at Dortmund, 44221 Dortmund, Germany}
\affiliation{Spin Optics Laboratory, St.~Petersburg State University, 198504 St.~Petersburg, Russia}
\author{I.~A.~Akimov}
\affiliation{Experimentelle Physik 2, Technische Universit\"at Dortmund, 44221 Dortmund, Germany}
\affiliation{Ioffe Institute, Russian Academy of Sciences, 194021 St.~Petersburg, Russia}
\author{V.~V.~Chaldyshev}
\affiliation{Ioffe Institute, Russian Academy of Sciences, 194021 St.~Petersburg, Russia}
\author{W.~V.~Lundin}
\affiliation{Ioffe Institute, Russian Academy of Sciences, 194021 St.~Petersburg, Russia}
\author{A.~V.~Sakharov}
\affiliation{Ioffe Institute, Russian Academy of Sciences, 194021 St.~Petersburg, Russia}
\author{A.~F.~Tsatsulnikov }
\affiliation{Submicron Heterostructures for Microelectronics, Research and Engineering Center, Russian Academy of Sciences, 194021, St.~Petersburg, Russia}
\author{D.~R.~Yakovlev}
\affiliation{Experimentelle Physik 2, Technische Universit\"at Dortmund, 44221 Dortmund, Germany}
\affiliation{Ioffe Institute, Russian Academy of Sciences, 194021 St.~Petersburg, Russia}
\author{M.~Bayer}
\affiliation{Experimentelle Physik 2, Technische Universit\"at Dortmund, 44221 Dortmund, Germany}
\affiliation{Ioffe Institute, Russian Academy of Sciences, 194021 St.~Petersburg, Russia}

\date{\today}

\begin{abstract}
We study the coherent dynamics of localized excitons in 100-periods of 2.5~nm thick (In,Ga)N/GaN quantum wells with 7.5\% indium concentration, measured with spectroscopic resolution through two-pulse and three-pulse photon echoes at the temperature of 1.5~K. A long-lived coherent exciton dynamics is observed in the (In,Ga)N quantum wells: When the laser photon energy is tuned across the 43~meV-wide inhomogeneously broadened resonance line, the coherence time $T_2$ varies between 45 and 255~ps, increasing with stronger exciton localization. The corresponding narrow homogeneous linewidths ranging from 5.2 to 29~$\mu$eV as well as the relatively weak exciton-phonon interaction ($0.8~\mu$eV/K) confirm a strong, quantum dot-like exciton localization in a static disordered potential inside the (In,Ga)N quantum well layers.
\end{abstract}

\maketitle

\section{Introduction}
\label{sec:1}

Compositional disorder in (In,Ga)N/GaN quantum wells (QWs) is an important phenomenon with strong impact on their optical and transport properties \cite{ChichibuAPL1996,NakamuraAPL1997,NarukawaAPL1997,SinghAPL1997}. It causes lateral in-plane localization of QW excitons \cite{ChichibuAPL1996,NakamuraAPL1997}, which affects the performance of (In,Ga)N/GaN QW based LEDs and lasers \cite{NarukawaAPL1997}. The disorder originates from the growth thermodynamics of the In-Ga-N alloy, in which the InN-GaN phase separation reduces the free energy of the system \cite{SinghAPL1997}.

In state-of-the-art growth methods, such as metal-organic vapor-phase epitaxy (MOVPE), the compositional disorder in (In,Ga)N/GaN QWs can be controlled by the growth conditions, which can lead to either a thermodynamically favorable highly non-uniform indium distribution or kinetically limited metastable atomic arrangements with almost statistical compositional fluctuations \cite{MusikhinAPL2002}.This makes possible to optimize the inhomogeneous broadening of the QW exciton transitions \cite{Bolshakov2015}. A reasonably large broadening seems to be in favor of a better performance of (In,Ga)N/GaN QW LEDs, whereas it has a negative impact on the performance of (In,Ga)N/GaN QW lasers \cite{Kawakami2001}.

The compositional disorder influences the exciton states in (In,Ga)N/GaN QWs in several ways. First, the corresponding inhomogeneous broadening spreads out and reduces the excitonic density of states, which is crucial for the gain in laser diodes \cite{Kawakami2001}. Second, it may destroy the coherence of excitons in multiple QW systems, which is the key condition for super-radiance of optical lattices of QW excitons \cite{ChaldyshevAPL2011}. Third, the compositional inhomogeneities create local electric fields, which cause a spatial separation of electrons and holes and reduce the exciton oscillator strength. Fourth, the spatial localization of excitons prevents them from inelastic scattering and recombination at structural defects such as dislocations. As a result of these factors, the excitons localized by compositional fluctuations in (In,Ga)N/GaN QWs are expected to have a long-term coherence due to reduced radiative and non-radiative decay and scattering rates. 

The most efficient tools for studying coherent exciton dynamics are two- and three-pulse photon echoes, which provide important information on the coherence time $T_2$ and the population decay time $T_1$, enabling additionally a precise measurement of the homogeneous linewidth. This is important and lacking information, since most studies of exciton localization in (In,Ga)N/GaN QWs have focused so far on the measurement of micro-photoluminescence (PL) lines from the individual localized exciton states at low temperatures \cite{SchoemigPRL2004, GotohAPL2006}. Only a few works are known where four-wave mixing techniques were applied to study the coherence of excitons in GaN bulk \cite{ZimmermannPRB1997, IshiguroPSSC2007}, epilayers \cite{SchneckAPL2012,Haag1997} and GaN-based QWs \cite{KundysPRB2006, GallartPRB2017}. In particular, exciton dephasing times as short as $T_2\sim300$~fs were measured by the photon echo technique in a 120~nm GaN film at temperature $T=10$~K \cite{SchneckAPL2012}. An even shorter exciton coherent dynamics was measured on 8-nm thick 10-period In$_{0.11}$Ga$_{0.89}$N/GaN QWs at 5~K using fs-pulses \cite{KundysPRB2006}. Sub-ps decay of exciton coherence was also observed recently in GaN/(Al,Ga)N QWs at low temperatures \cite{GallartPRB2017}.

In this paper, we study the coherent dynamics of excitons localized by compositional fluctuations in (In,Ga)N/GaN QWs using the time-resolved photon echo technique. We use 100 QWs in order to enhance the coherent response from the QW excitons. Applying two- and three-pulse excitation protocols we observe the primary and stimulated photon echoes, which deliver rich information on the coherent evolution of the QW excitons. We employ spectrally tunable picosecond laser pulses to address different localized exciton states and obtain the low-temperature homogeneous linewidth of the exciton from the temporal decay of the two-pulse photon echo. We reveal a long-term coherent dynamics of the excitons reaching up to 255~ps. We also see that the acoustic phonons rapidly destroy the exciton coherence at elevated temperatures.

\begin{figure}[t]
	\vspace{5mm}
	\includegraphics[width=\linewidth]{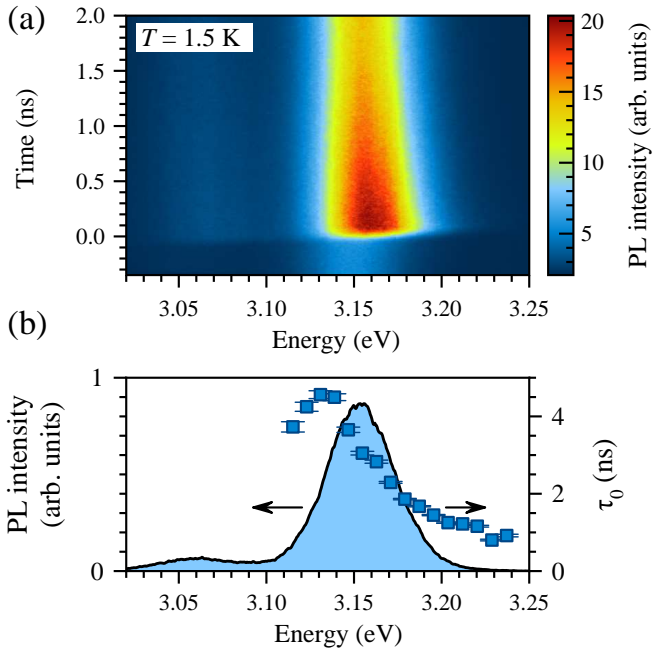}
	\caption{(a) TRPL signal detected with the streak camera at $T=1.5$~K. Excitation pulse photon energy is 3.446~eV. (b) Spectral dependences of exciton lifetime $\uptau_0$ (squares) extracted from the TRPL measurements across the PL emission spectrum (line).}
	\label{TRPL}
\end{figure}

\section{Sample and Method}
\label{sec:2}

The  sample was grown by metal-organic vapor-phase epitaxy (MOVPE) on a GaN-on-sapphire substrate with a 2-$\mu$m-thick GaN layer. A 700 nm GaN buffer was grown at a temperature of 1030$^\circ$C under a pressure of 200~mbar. It was followed by 100 periods of 2.5~nm-thick (In,Ga)N QWs separated by GaN barriers. The wafer was not rotated during the growth procedure and, therefore, the sample has an intentional gradient of layer thickness and composition resulting in a variation of the absorption and PL spectra across the sample. From the mapping and simulations of high-resolution X-ray diffraction curves described in Ref. \cite{ChaldyshevJAP2017}, the averaged indium concentration and QW width are estimated to be 7.5\% and 2.5~nm, respectively, and the measured period of the multiple QW system is about 72~nm, leading to 69.5~nm barrier width. As a result of the periodicity we observed a Bragg reflection peak in the stationary optical reflection spectra \cite{Bolshakov2015}. At cryogenic temperatures the peak is centered at 3.393~eV (365.4~nm) and its full width is about 32~meV (3.5~nm). In this work we study excitons at substantially lower energies far from the Bragg resonance. Thereby, we do not expect prominent effects connected to exciton-light interaction changes due to the QW periodicity.

In order to identify the spectral position of the exciton emission and measure the exciton lifetime, time-resolved photoluminescence (TRPL) with spectral resolution was measured in transmission at $T=1.5$~K, using a streak camera. Excitation was done by picosecond laser pulses tuned to the energy of 3.446~eV (360~nm). The TRPL dynamics recorded with  20~ps temporal resolution is shown in Fig.~\ref{TRPL}(a). It demonstrates mainly a broad spectral line with different decay times varying considerably across the emission. The time-integrated PL spectrum, shown in Fig.~\ref{TRPL}(b), reveals two broad lines with Gaussian shapes. The main line with full width at half maximum (FWHM) of 46~meV is located at 3.154~eV, while the phonon replica with FWHM 62~meV is red-shifted by 89.5~meV, corresponding to the optical phonon energy \cite{Nakamura1996}. The PL dynamics is analyzed using exponential fits of the PL decays at different energies. The obtained spectral dependence of the exciton lifetime $\uptau_0$, plotted in Fig.~\ref{TRPL}(b), displays an almost monotonous increase from 0.8 up to 4.5~ns with decreasing exciton emission energy.

\begin{figure}[t]
	\vspace{5mm}
	\includegraphics[width=\linewidth]{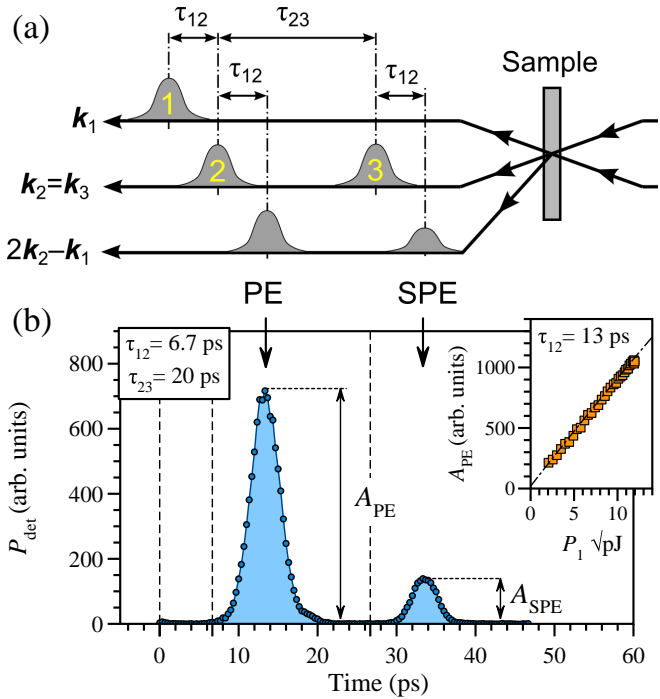}
	\caption{Photon echo detection: (a) Optical scheme of sample excitation in transmission geometry with timing of excitation pulses and echo signals. PE and SPE denote two-pulse primary and three-pulse stimulated photon echoes, respectively. (b) Experimental FWM transient consisting of PE and SPE peaks. Dashed lines indicate temporal positions of the three exciting pulses separated by $\tau_{12}=6.7$~ps and $\tau_{23}=20$~ps. Inset shows PE amplitude $A_{\text{PE}}$ as a function of first pulse amplitude $P_1$ calculated as square root of pulse energy, in combination with a  linear fit.}
	\label{setup}
\end{figure}

In order to study the coherent evolution and population dynamics of excitons, four-wave mixing (FWM) with picosecond temporal resolution was realized, allowing generation and detection of two-pulse and three-pulse photon echoes. The sample was immersed in liquid helium inside the variable temperature insert of an optical bath cryostat. Optical excitation of the sample was performed by a sequence of picosecond laser pulses generated by a tunable Ti:Sapphire laser operating at a repetition rate of 75.8 MHz. The optical frequency of the pulses was doubled by a second-harmonic generator. The first and second pulses, separated by the variable temporal delay $\tau_{12}$, were focused into a spot of about 300~$\mu$m in diameter, hitting the sample almost along its normal having the wave vectors $\bm{k}_1$ and $\bm{k}_2$ with an angular difference of about 1$^\circ$ between them. The third pulse with $\bm{k}_3=\bm{k}_2$ was delayed by the time interval $\tau_{23}$ relative to the second pulse. All pulses were linearly co-polarized. The FWM signal was collected in the transmission geometry along the direction $2\bm{k}_2-\bm{k}_1$ as shown in Fig.~\ref{setup}(a). In order to detect a background-free FWM signal with temporal resolution we exploited optical heterodyning and interferometric detection, using an additional reference laser pulse as described elsewhere \cite{PoltavtsevZnO2017}. The detected signal is the absolute value of the cross-correlation of the FWM field with the reference pulse field.

\section{FWM Experimental results}
\label{sec:3}

The excitation of the (In,Ga)N/GaN QWs by resonant laser pulses results in a robust FWM signal in form of photon echoes. A typical FWM transient after application of the three pulses with delays $\tau_{12}=6.7$~ps and $\tau_{23}=20$~ps is demonstrated in Fig.~\ref{setup}(b). It consists of the primary photon echo (PE) pulse located at $2\tau_{12}=13.3$~ps and the stimulated photon echo (SPE) pulse located at $2\tau_{12}+\tau_{23}=33.3$~ps. The temporal profile of the PE pulse has a Gaussian shape with FWHM of 4.3~ps. This corresponds to the convolution of the echo pulse from the excited sub-ensemble with the Fourier transform limited laser pulse with duration (FWHM in field domain) $\tau_P=1.8$~ps: $\sqrt{6}\tau_P\approx4.4$~ps. The amplitude of the PE signal $A_{\text{PE}}$ was checked to be linearly dependent on the first pulse amplitude (square root of the pulse energy), as shown in the inset of Fig.~\ref{setup}(b), up to the the maximal available optical power density of 230~nJ/cm$^2$ provided by the first pulse, while the second and third pulses were at least twice less intense. Thereby, all experiments presented here were conducted in the $\chi^{(3)}$ regime.

\begin{figure*}[t]
	\vspace{5mm}
	\includegraphics[width=\linewidth]{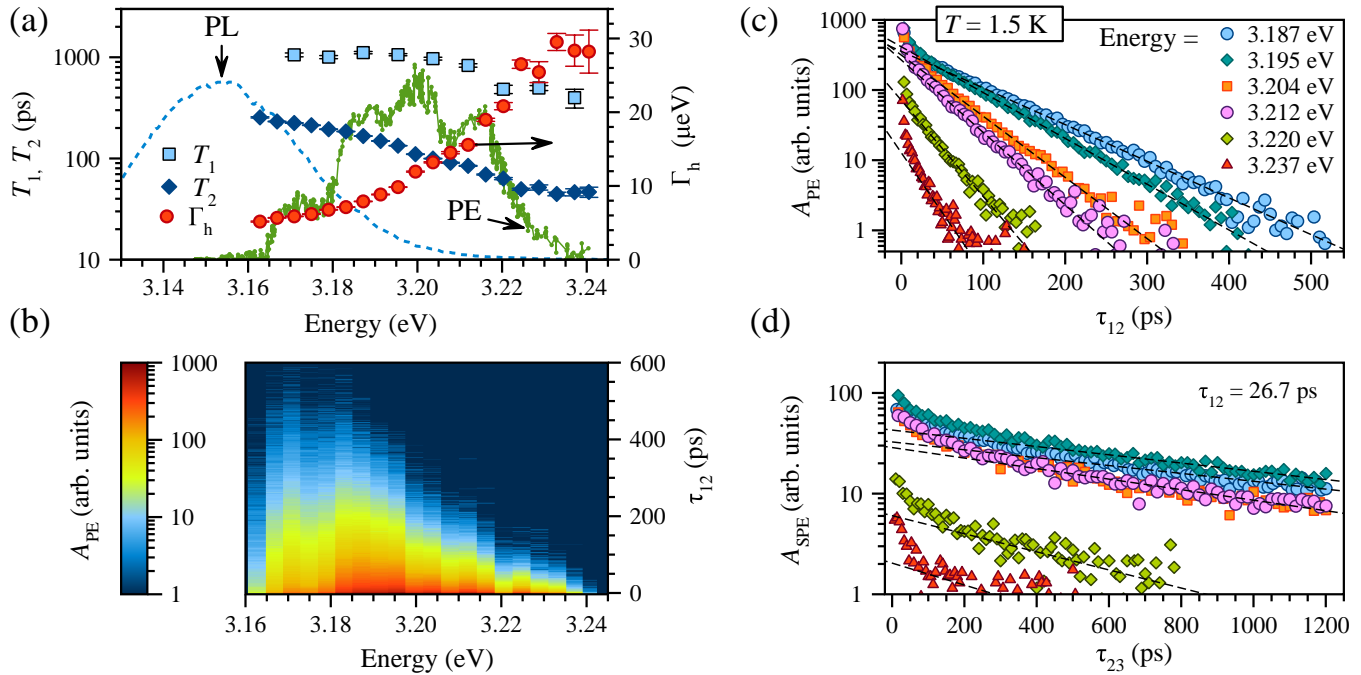}
	\caption{Two-pulse and three-pulse photon echo spectroscopy accomplished at $T=1.5$~K: (a) Spectral dependences of population decay time $T_1$, coherence time $T_2$ and homogeneous linewidth $\Gamma_\text{h}=2\hbar/T_2$, extracted from the PE and SPE decays. Blue dashed-line profile on the low-energy side corresponds to the PL spectrum. Green solid-line profile on the high-energy side corresponds to the PE amplitude spectrum, detected at the pulse delay $\tau_{12}=13.3$~ps. (b) Spectral dependence of PE decay measured with a step of 4~meV. (c) PE and (d) SPE decays measured for a variety of laser energies (different points) with exponential fittings (dashed lines). SPE is detected at $\tau_{12}=26.7$~ps.}
	\label{PE_SPE}
\end{figure*}

The narrow pulse spectral width of 1.7~meV and the large tuning range 2.5--3.5~eV (350--495~nm) made it possible to address exciton transitions with different localization energies and, thus, to study photon echoes with spectroscopic resolution. Figure~\ref{PE_SPE} is the summary of the two-pulse and three-pulse photon echo spectroscopy accomplished on the studied (In,Ga)N/GaN QWs at $T=1.5$~K. Surprisingly, echo signal was found in a fairly broad spectral range, covering laser energies $E=$3.16--3.24~eV (382--392~nm). By varying the pulse photon energy at fixed time delay $\tau_{12}=13.3$~ps we were able to record the two-pulse echo spectrum, displayed in Fig.~\ref{PE_SPE}(a) and located on the high-energy side. It can be well fitted by a Gaussian shape with FWHM $\Gamma_2^*=43$~meV ($\sim5$~nm) with a periodic modulation presumably emerging from the Fabry-P\'erot interference between the sample surface and the interface of the structure with sapphire. Such a huge inhomogeneous broadening compared to the pulse spectral width (1.7~meV) ensures that the photon echo is always generated by a sub-ensemble selected by the pulse spectrum, in full accord with the echo's temporal profile [Fig.~\ref{setup}(b)]. Interestingly, the broad PL spectrum, located in Fig.~\ref{PE_SPE}(a) on the low-energy side, weakly overlaps with the PE spectrum, demonstrating an effective Stokes shift of 46~meV calculated as the difference between the PL and PE line centers. Since both signals were measured in transmission, the PL spectrum may be in effect red-shifted due to the QW absorption. A Stokes shift as large as 20~meV was observed earlier  at room temperature by comparison of the optical absorption and PL spectra for similar samples \cite{ChaldyshevJAP2017}.

By variation of the $\tau_{12}$ pulse delay and detecting the PE pulse amplitude $A_{\text{PE}}$ at fixed laser energy we measured the PE decay, which allows determination of the exciton coherence time $T_2$: $A_{\text{PE}}\sim\exp(-2\tau_{12}/T_2)$ \cite{BermanMalinovskyBook}. PE decay measurements for different laser energies within the PE spectrum result in the two-dimensional $A_{\text{PE}}(\tau_{12},E)$ dependence plotted in Fig.~\ref{PE_SPE}(b). The PE decays plotted in Fig.~\ref{PE_SPE}(c) for a variety of selected laser energies demonstrate almost pure single exponential decays, from which we obtain a spectral dependence of coherence times $T_2$ varying monotonously between 45 and 255~ps, when the laser photon energy is tuned across the wide exciton line. This corresponds to a variation of the homogeneous linewidth $\Gamma_\text{h}=2\hbar/T_2$ between 5.2 and 29~$\mu$eV, as shown in Fig.~\ref{PE_SPE}(a) with circles.

In the three-pulse experiment, $\tau_{12}$ delay is fixed and variation of the $\tau_{23}$ delay provides measurement of the population decay time $T_1$: $A_{\text{SPE}}\sim\exp(-\tau_{23}/T_1)$. The experimental SPE decays measured at $\tau_{12}=13.3$~ps and plotted in Fig.~\ref{PE_SPE}(d) manifest a non-trivial dynamics reflecting the presence of exciton states with two different decay times. From fitting of the slow decaying component of these dynamics, we can extract the $T_1$ plotted in Fig.~\ref{PE_SPE}(a) with squares. It varies between 0.4 and 1.1~ns, which is much longer than the exciton coherence time $T_2$ and rather close to the exciton lifetime $\uptau_0$ that varies in the range from 0.9 to 2.3~ns in the same energy range. The fast decaying component is difficult to analyze, however, short $T_1$ times can be estimated to be in the range 50--100~ps, which is much closer to the obtained $T_2$ values.

\begin{figure}[t]
	\vspace{5mm}
	\includegraphics[width=\linewidth]{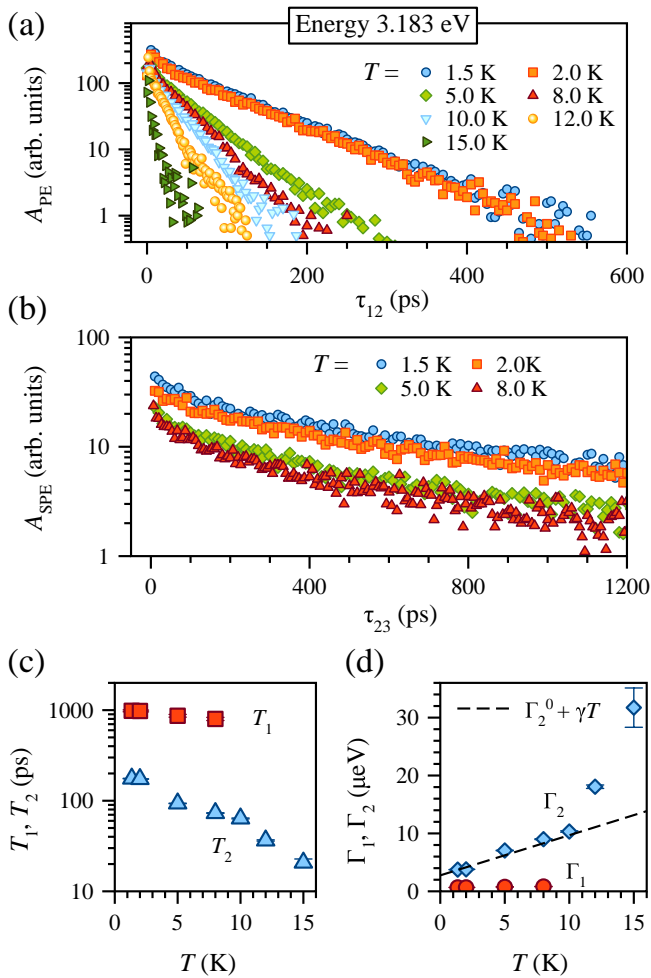}
	\caption{Temperature dependence of (a) two-pulse and (b) three-pulse photon echo decays measured at energy 3.183~eV. (c) Extracted dependences of (c) decay times $T_2$, $T_1$ and (d) decay rates $\Gamma_2$, $\Gamma_1$ on temperature. A linear fit of the decoherence rate at temperatures up to 10~K is shown by dashed line. Population decay rate is constant within $\Gamma_1=0.7\pm0.1$~$\mu$eV.}
	\label{PE_SPE(T)}
\end{figure}

The role of acoustic phonons can be understood through measurements of the photon echo decays at varied temperatures. The results of PE and SPE decay measurements performed at the laser energy of 3.183~eV are summarized in Fig.~\ref{PE_SPE(T)}. While the two-pulse photon echo decays experience a strong shortening with rising temperature, the three-pulse photon echo manifests mainly a decrease of the $A_{\text{SPE}}$ amplitude, demonstrating an almost constant decay rate. The temperature dependences of the $T_2$, $T_1$ times and the $\Gamma_2$, $\Gamma_1$ decay rates are displayed in Figs.~\ref{PE_SPE(T)}(c) and \ref{PE_SPE(T)}(d), accordingly. The low-temperature part of the temperature dependence of the decoherence rate can be approximated with a linear dependence $\Gamma_2=\Gamma_2^0+\gamma T$. The slope $\gamma=0.8~\mu$eV/K responsible for the exciton dephasing due to interaction with acoustic phonons is somewhat smaller than that observed recently in ZnO bulk and QWs ($\sim2~\mu$eV/K) \cite{PoltavtsevZnO2017, SolovevPRB2018}. It can be concluded, however, that acoustic phonons strongly and unavoidably damp exciton coherence in (In,Ga)N/GaN QWs.

\section{Discussion}
\label{sec:4}

The spectral behavior of the coherence time $T_2$ and the population decay time $T_1$ can be explained by the localization of excitons on the QW potential fluctuations. Qualitatively similar $T_2$ and $T_1$ spectra measured through photon echoes were acquired recently on negatively charged excitons (trions) in ZnO QWs: Both spectral dependences revealed at $T=1.5$~K a monotonous increase with decreasing energy and the population decay lasted substantially longer than the trion dephasing ($T_1\gg T_2/2$) \cite{SolovevPRB2018}. 

It was already noted above that the compositional disorder in (In,Ga)N/GaN QWs is strong with details depending on technology. The sample under study was grown with special care to reduce the compositional broadening \cite{Bolshakov2015}. Nevertheless, the Stokes shift as large as 20 meV at room temperature \cite{ChaldyshevJAP2017} indicates a strong lateral localization of the QW excitons. Origin and scale of the potential fluctuations in (In,Ga)N/GaN QWs were investigated by various methods, including transmission electron microscopy \cite{MusikhinAPL2002}, near-field scanning optical microscopy \cite{Kim2017}, and scanning tunneling luminescence spectroscopy \cite{HahnPRB2018}. In particular, Hahn et al. showed that micro-PL spectra display significant and chaotic potential variations on the scale of about 3~nm \cite{HahnPRB2018}. This is commensurable with the free exciton Bohr radius in GaN of about 3~nm \cite{Chichibu1999} and might provide a reason for the modification of the $T_1$ time connected with the exciton oscillator strength. The fact that we observe a somewhat weaker exciton interaction with the acoustic phonons ($0.8~\mu$eV/K) than in QWs of other types additionally points towards the quantum dot-like exciton localization in (In,Ga)N/GaN QWs. In InGaAs/GaAs quantum dots this interaction was measured to be  $\gamma=0.5~\mu$eV/K \cite{BayerPRB2002} or was even practically absent \cite{BorriPRB2005}.

Another potential reason for the observed three-fold shortening of the population decay with rising exciton energy is the exciton-light interaction enhancement due to the spatial periodicity of the studied QWs. If the Bragg period of the light waves allows Bragg diffraction at the frequency of the QW excitons, then a collective superradiant exciton-polariton mode is formed due to electromagnetic coupling of all $N$ QWs and the radiative decay rate becomes $N$-times enhanced \cite{Ivchenko1994}. Shortening of the coherent response decay was indeed observed previously in FWM experiments on GaAs-based resonant Bragg structures \cite{HuebnerPRL1996}. Here we detect photon echoes at energies essentially lower (about 6\%) than the Bragg peak location (3.392~eV), so that such effects are hardly expected. Moreover, the huge difference between the exciton homogeneous linewidth of about 20~$\mu$eV and the inhomogeneous broadening of 43~meV makes radiative coupling between different QWs at low temperatures improbable.

Since coherent dynamics of localized excitons in (In,Ga)N/GaN QWs in picosecond range has not been reported so far it is difficult to compare directly our findings with other studies. Most of the works studying strongly localized excitons analyze micro-PL spectra measured from individual states at low temperatures. PL spectra measured from a 3~nm-thick In$_{0.15}$Ga$_{0.85}$N/GaN single QW at $T=4$~K demonstrate linewidths of individual exciton states down to 0.8~meV \cite{SchoemigPRL2004}. PL linewidths down to 0.4~meV were measured at a similar temperature in In$_{0.1}$Ga$_{0.9}$N/GaN multiple QWs, in which static potential fluctuations were laterally isolated using 0.2~$\mu$m wide mesas \cite{GotohAPL2006}. The authors of this study concluded about the quantum dot-like character of the potential minima, formed by spatial indium accumulation. To make a comparison with our results, we estimate the homogeneous linewidth at $T=4$~K as 11--35~$\mu$eV depending on the exciton energy. These estimates take into account the energy-independent line broadening due to the exciton-phonon interaction, $2\gamma T\approx6~\mu$eV. These estimates are still at least one order of magnitude smaller than the PL linewidths observed from the individual exciton states. Such a large difference between the exciton homogeneous linewidths estimated here and the previously observed PL linewidths might be caused by an effective micro-PL line broadening due to spectral jitter during exposure, or dynamical broadening, e.g. due to charge fluctuations. Because of the short pulse delays the photon echoes are insensitive to such jitter processes.

\section{Conclusions}
\label{sec:5}

Photon echoes in the violet spectral range were observed from localized excitons in multiple (In,Ga)N/GaN quantum wells. Using two-pulse resonant excitation with picosecond laser pulses we were able to measure the homogeneous linewidth of the localized excitons at cryogenic temperature with spectroscopic resolution. We found that the localized excitons in our MQW structure show a much longer coherent dynamics than it could be expected from previous micro-PL studies of (In,Ga)N/GaN QWs with higher indium concentrations. The exciton coherent life lasts up to 255~ps, which is substantially longer than typical coherence times of excitons localized in III-V QWs composed of GaAs/(Al,Ga)As \cite{CundiffPRB1992}, (In,Ga)As/GaAs \cite{PoltavtsevSSC2014}, or in II-IV QWs such as CdTe/(Cd,Mg)Te \cite{PoltavtsevCdTe2017} and ZnO/(Zn,Mg)O \cite{SolovevPRB2018}, where the typical coherence times are in the range of tens of ps. We also see that across the wide spectrum of localized  states there is a strong correlation between the exciton localization energy and its coherence times, which differ by a factor of five between the lower and higher energy flanks of this line. 

{\it Acknowledgments.} The authors thank S. V. Goupalov for helpful discussions. The authors acknowledge the Deutsche Forschungsgemeinschaft (DFG) for financial support through the Collaborative Research Centre TRR 142 (Project No. A02) and the International Collaborative Research Centre 160 (Project No. A3). I.A.S. and S.V.P. thank the Russian Foundation for Basic Research (RFBR) for partial financial support of their work (Research Grant No. 15-52-12016 NNIO\_a) and acknowledge St. Petersburg State University for the Research Grant No. 11.34.2.2012. I.A.S. thanks RFBR for the Research Grant No. 18-32-00684 mol\_a.


\end{document}